\documentclass[aps,showpacs,prl,
reprint,
  ]{revtex4-1}
\usepackage{graphicx}
\usepackage{natbib}
\usepackage{amsmath}
\usepackage{epstopdf}
\usepackage{color}
\usepackage{upgreek}
\usepackage{gensymb}
\begin{document}

\title{A controllable two-membrane-in-the-middle cavity optomechanical system}
\author{Xinrui Wei}
\author{Jiteng Sheng$^{*}$}
\author{Cheng Yang}
\author{Yuelong Wu}
\author{Haibin Wu$^{\dag}$}

\affiliation{State Key Laboratory of Precision Spectroscopy, East China Normal University, Shanghai 200062, China}


\begin{abstract}
We report an optomechanical system with two dielectric membranes inside a Fabry-Perot cavity. The cavity resonant frequencies are measured in such a two-membrane-in-the-middle system, which show an interesting band-structure-like diagram. This system exhibits great controllability on the parameters of the system. The positions and angles of each membrane can be manipulated on demand by placing two membranes inside the cavity separately. The eigenfrequencies of the vibrational modes of the membranes can also be tuned individually with piezoelectricity. This scheme could be straightforwardly extended to multiple-membrane-in-the-middle systems, where more than two membranes are involved. Such a well controllable multiple membrane optomechanical system provides a promising platform for studying nonlinear and quantum dynamical phenomena in multimode optomechanics with distinct mechanical oscillators.
\end{abstract}

\maketitle

Recently, cavity optomechanics has attracted considerable attention due to its wide applications in highly sensitive measurements, quantum information, quantum computing, nanophotonics, as well as the fundamental test of quantum mechanics in macroscopic systems \cite{Kippenberg1172,RevModPhys.86.1391,1742-6596-264-1-012025}. The minimal model of cavity optomechanics considers the coupling between one optical mode and one mechanical mode. Multimode optomechanics, where two or more optical or mechanical modes are involved, provides an opportunity to reach some regimes where the minimal model is difficult to achieve and to study some more rich and complicated physical phenomena, such as enhanced optomechanical coupling, collective dynamics, mechanical motion entanglement, energy transfer, and synchronization \cite{PhysRevLett.107.043603,1367-2630-10-9-095009,Lin2010,PhysRevLett.111.073603,Nielsen62,Massel2012,PhysRevLett.112.013602,nature537,naturecomm2017,PhysRevLett.111.213902,Li2014,PhysRevLett.109.233906,PhysRevLett.118.063605}. Among various multimode optomechanical schemes, an important model is a Fabry-Perot cavity with multiple dielectric membranes in the middle. Although this model has been extensively explored in various theoretical proposals \cite{PhysRevA.78.041801,PhysRevLett.101.200503,PhysRevLett.109.223601,PhysRevA.86.063829,PhysRevA.86.033821,PhysRevA.90.053808,PhysRevA.92.013851,PhysRevA.97.022336}, experimental attempts in this regard just begin recently \cite{Nair:17,1367-2630-20-8-083024}. 

Very recently, the first experimental realization of two-membrane cavity optomechanics is reported by Piergentili et al \cite{1367-2630-20-8-083024}. Some important parameters of the system are difficult to tune in their setup. Here, we realize a well controllable two-membrane cavity optomechanical system. The membranes are inserted into the cavity separately. This system provides several advantages. Firstly, membranes can be placed at any position, in principle, inside the cavity. This could be very important when the membranes are required to be placed inside the cavity asymmetrically \cite{PhysRevLett.101.200503}. Moreover, the angle of each membrane can be tilted independently, which is necessary to eliminate or reduce high-order cavity modes \cite{Lee2015}. In addition, the eigenfrequencies of the vibrational modes of the membranes can be tuned on demand with piezoelectricity \cite{doi:10.1063/1.5009952}. As a result, the frequency difference between mechanical modes of each membrane can be set to different values in a wide parameter region, which will be very useful to study synchronization of multiple optomechanical membranes \cite{pikovsky2003synchronization,PhysRevA.96.023805}. 

In this work, we mainly investigate the band-structure-like diagrams, or cavity resonant frequency as a function of two-membrane positions, in such a two-membrane-in-the-middle optomechanical system. The controllability of the cavity resonant frequencies is demonstrated. The band structure is an essential property for a membrane-in-the-middle system, since it reveals one of the key parameters, i.e. single photon coupling coefficient. The situation becomes more complicated in our case when two membranes inside an optical cavity are involved. Not only can the single photon coupling coefficient be enhanced \cite{1367-2630-20-8-083024}, but also the band structure turns into three-dimensional (3D). The bands for various collective modes are completely different compared with the single membrane case and there exist dark modes where the bands are flat. Therefore, a comprehensive study of the band structure is a prerequisite for further investigations of such a two-membrane-in-the-middle optomechanical system.

\begin{figure}
\centering
\includegraphics[width=1\linewidth]{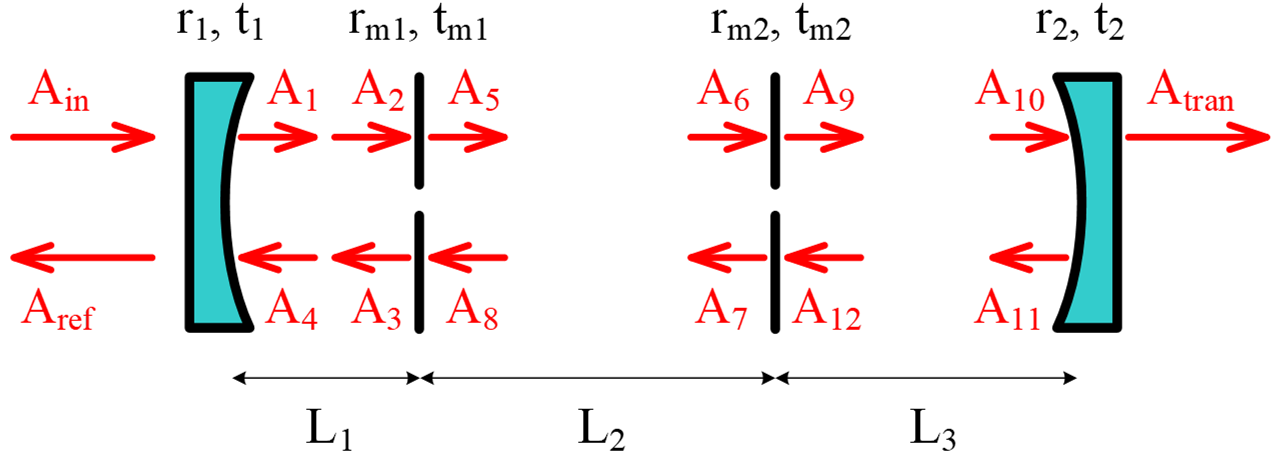}
\caption{Schematic diagram of the two-membrane-in-the-middle cavity optomechanical system. Two membranes divide the Fabry-Perot cavity into three coupled cavities, which have lengths of $L_1=30$, $L_2=60$, and $L_3=50$ $mm$, respectively.}
\end{figure}

A schematic of our experimental system is shown in Fig. 1. The optical cavity consists of two identical mirrors, which has a length of 140 $mm$ and a finesse $\sim$ 1000. Two stoichiometric silicon nitride (SiN) membranes with a thickness of 50 $nm$ and a $1\times1$ $mm^2$ size are placed inside the cavity asymmetrically, as depicted in Fig. 1. Thus, the two mirrors and two membranes form three coupled cavities, which have lengths of $L_1$, $L_2$, and $L_3$, respectively. The transmission and reflection of the cavity with membranes in the middle are quite different from those of the bare two-mirror Fabry-Perot cavity. Such a multiple-membrane-in-the-middle system is very similar to N-coupled ring resonators \cite{Smith:03}. Specifically, the system with two membranes is identical to three-coupled ring resonators \cite{doi:10.1002/lpor.201600178}. Hence, the iterative approach \cite{Smith:03} for analysis of coupled ring resonators should be feasible to such a multiple-membrane-in-the-middle system. Other methods to calculate cavity resonance frequencies are the Helmholtz equation with boundary conditions \cite{PhysRevA.78.041801}, transfer matrix \cite{PhysRevA.94.053812}, amplitude equations \cite{1367-2630-10-9-095008,1367-2630-20-8-083024}, et al.  

Here we adopt the method of amplitude equations. By denoting $A_i$ as the electric field amplitudes ($i=1,2,...,10$ are the intracavity fields, $i=in$ is the cavity input, $i=ref$ is the cavity reflection, and $i=tran$ is the cavity transmission), we can obtain the following equations
\begin{align}
 A_1  = it_1 A_{in}  + r_1 A_4,  \label{eqn:1}\\ 
 A_2  = A_1 e^{ - ikL_1 },  \label{eqn:2}\\ 
 A_3  = it_{m1} A_8  + r_{m1} A_2,  \label{eqn:3}\\ 
 A_4  = A_3 e^{ - ikL_1 },  \label{eqn:4}\\ 
 A_5  = it_{m1} A_2  + r_{m1} A_8,  \\ 
 A_6  = A_5 e^{ - ikL_2 },  \\ 
 A_7  = r_{m2} A_6  + it_{m2} A_{12},  \\ 
 A_8  = A_7 e^{ - ikL_2 },  \\ 
 A_9  = it_{m2} A_6  + r_{m2} A_{12},  \\ 
 A_{10}  = A_9 e^{ - ikL_3 },  \\ 
 A_{11}  = r_2 A_{10}  \\ 
 A_{12}  = A_{11} e^{ - ikL_3 },  \\ 
 A_{ref}  = it_1 A_4  + r_1 A_{in},  \\ 
 A_{tran}  = it_2 A_{10},
\end{align}
where $k=2\pi/\lambda$ is the angular wavenumber of light field, $r_i$ and $t_i$ ($i=1,2,m1,m2$) are the amplitude reflection and transmission coefficients of the mirrors and membranes, respectively. Here we assume that the membranes are thin and the phase shift due to the membrane is ignored \cite{PhysRevA.94.053812}.

By solving Eqs. (1-14) numerically, we can obtain a 3D band-structure-like diagram for the transmission of the two-membrane-in-the-middle system. As shown in Fig. 2, the vertical axis presents the cavity resonance frequency shift, which can be analogous to the energy of the electronic band structure. While the cavity resonance frequency shift depends on the positions of two membranes, instead of the wave vectors in the electronic band diagram. In other words, two membranes form a two-dimensional (2D) parameter space, which determines the resonance frequencies or bands of the coupled cavity system. Figures 2(a) and 2(b) represent the theoretical simulation where the two membranes are symmetrically and asymmetrically distributed inside the cavity, respectively. Only four out of many bands are illustrated in Figs. 2(a) and 2(b). As one can see in Fig. 2(a), the bottom and top bands are the same, which implies that the period of the band structure is three free-spectral-range (FSR) of the empty cavity in the symmetric case. In Fig. 2(b), such periodicity disappears, while the bottom and top bands are still the same, except one of them shifts horizontally with respect to the other. This property will be illustrated more clearly in 2D diagrams when one of the membrane's position is fixed, as will be discussed in the following. The amplitude transmission coefficient of the membrane is chosen to be 0.9, which is determined by the refractive index of SiN and the wavelength of light field \cite{1367-2630-10-9-095008,1367-2630-20-8-083024}.

\begin{figure}
\centering
\includegraphics[width=1\linewidth]{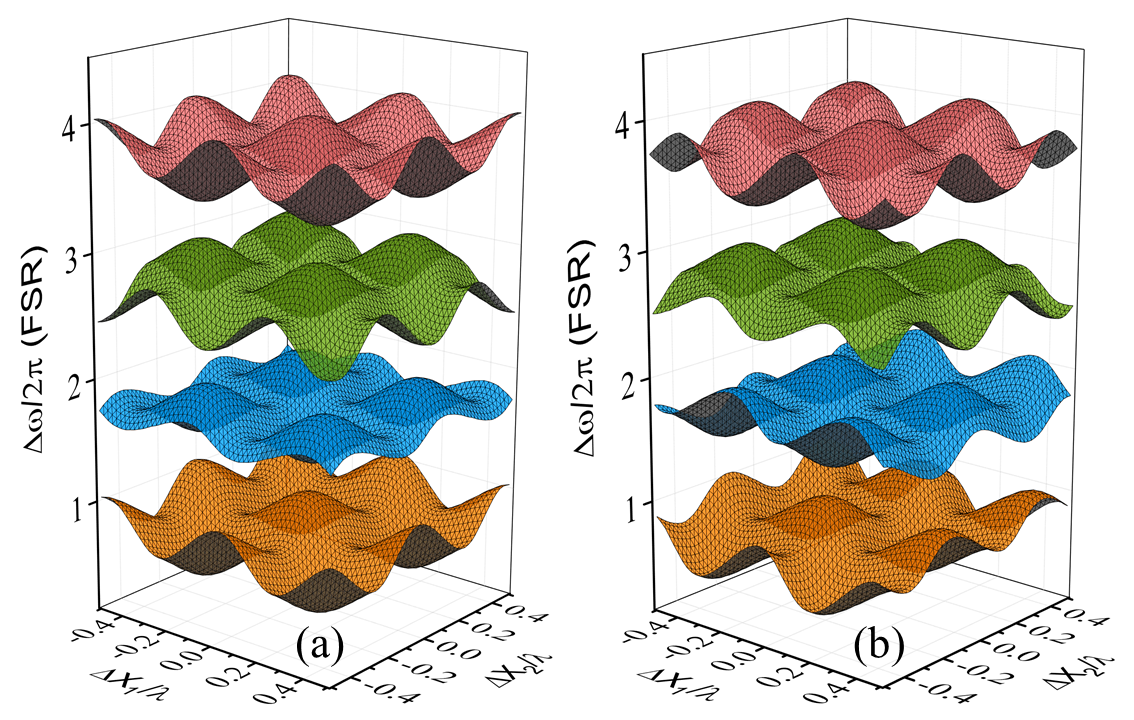}
\caption{Calculated band-structure-like diagram for the two-membrane-in-the-middle system. $\Delta\omega$ is the cavity resonance frequency shift. $\Delta x_1$ and $\Delta x_2$ are the position changes of two membranes. As the membranes move, the lengths of three coupled cavity become $L_1+\Delta x_1$, $L_2-\Delta x_1+\Delta x_2$, and $L_3-\Delta x_2$, respectively. (a) and (b) are the symmetric and asymmetric cases, where $L_1=L_2=L_3=47$ $mm$ for (a) and $L_1=30$, $L_2=60$, and $L_3=50$ $mm$ for (b), respectively. Other parameters for the calculation are $t_{m1}=t_{m2}=0.9$. The losses on the mirrors and membranes are ignored.}
\end{figure}

To better understand the cavity resonance frequency shift as a function of two-membrane positions, we investigate 2D band structure in four different cases, i.e. (1) $\Delta x_2$ is fixed and $\Delta x_1$ changes, (2) $\Delta x_1$ is fixed and $\Delta x_2$ changes, (3) both $\Delta x_1$ and $\Delta x_2$ change, and they satisfy $\Delta x=\Delta x_1=\Delta x_2$, and (4) $\Delta x=\Delta x_1=-\Delta x_2$. These four cases correspond to the projections of the 3D band-structure on a plane of (1) perpendicular to the axis of $\Delta x_2$, (2) perpendicular to the axis of $\Delta x_1$, (3) 45$^\circ$ relative to the axis of $\Delta x_1$, and (4) -45$^\circ$ relative to the axis of $\Delta x_1$. It is worth mentioning that the cases (1) and (2) reduce to single membrane systems, while the cases (3) and (4) are related to the center-of-mass mode and breathing mode \cite{PhysRevLett.109.223601}, respectively. 

\begin{figure*}
\centering
\includegraphics[width=1\linewidth]{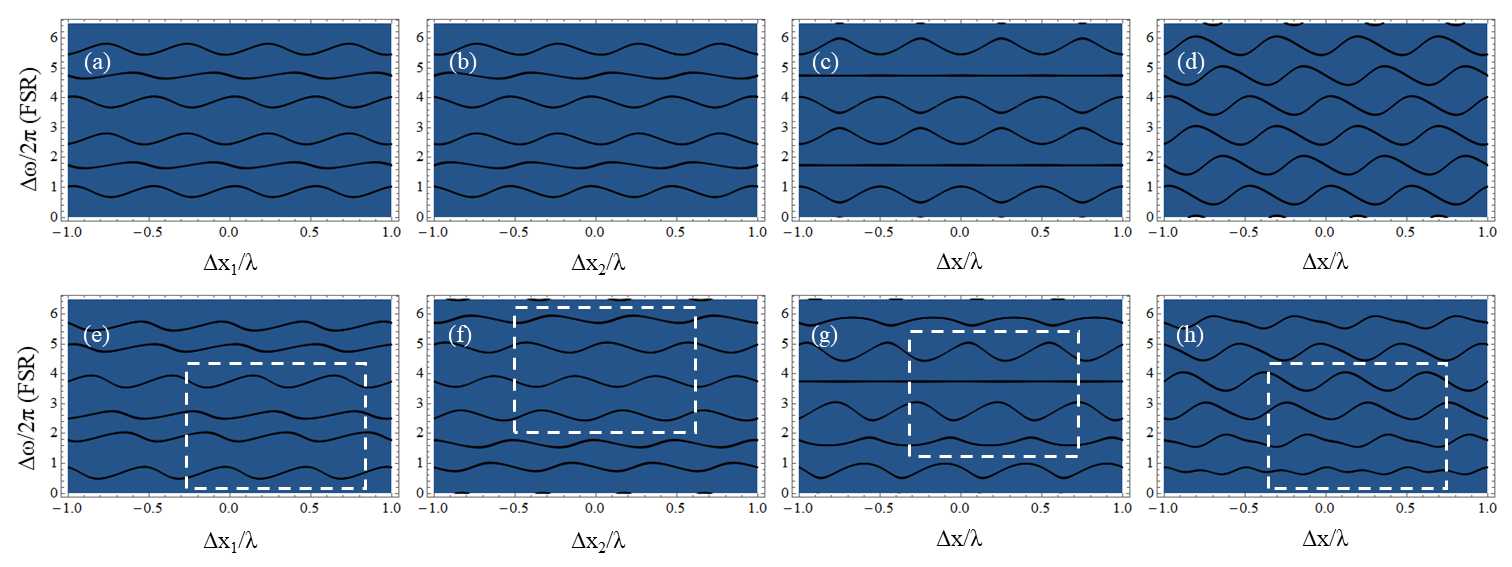}
\caption{Cavity resonance frequency shift as a function of two-membrane positions regarding the four different cases. (a-d) and (e-f) are the symmetric and asymmetric cases.}
\end{figure*}

Figures 3 shows the theoretical calculation of the cavity resonance frequency shift as a function of two-membrane positions regarding the four cases mentioned above. The symmetric and asymmetric cases are shown in Figs. 3(a)-3(d) and Figs. 3(e)-3(h), which correspond to the situations shown in Figs. 2(a) and 2(b), respectively. The periodicity of the symmetric case can be clearly seen in Figs. 3(a) and 3(b), while there is no such periodicity in the asymmetric case. In Figs. 3(c) and 3(g), the distance between two membranes are constant, and one can find that some bands are flat, which indicates that these collective modes are dark modes, i.e., no coupling to the cavity field. Please note that the flat band corresponds to a global and complete dark mode, which means that the coupling coefficient is always zero at any position on the band and the high-order coupling is also zero. This is different compared to the local dark mode \cite{1367-2630-10-9-095008}, where the coupling is zero only at the nodes and the quadratic coupling is not zero. Regarding the single membrane cases, i.e., Figs. 3(a), 3(b), 3(e), and 3(f), there is no flat band no matter where the membranes are positioned. The single photon optomechanical coupling coefficient is proportional to $\Delta\omega/\Delta x_{1,2}$, which is exactly the slope of the curves in Fig. 3. Therefore, by choosing different membrane positions and collective modes, the optomechanical coupling can be widely tuned. Moreover, the single photon coupling coefficient of certain collective mode is larger than that of the single membrane mode. This result is consistent with previous studies \cite{1367-2630-20-8-083024}.

\begin{figure}
\centering
\includegraphics[width=1\linewidth]{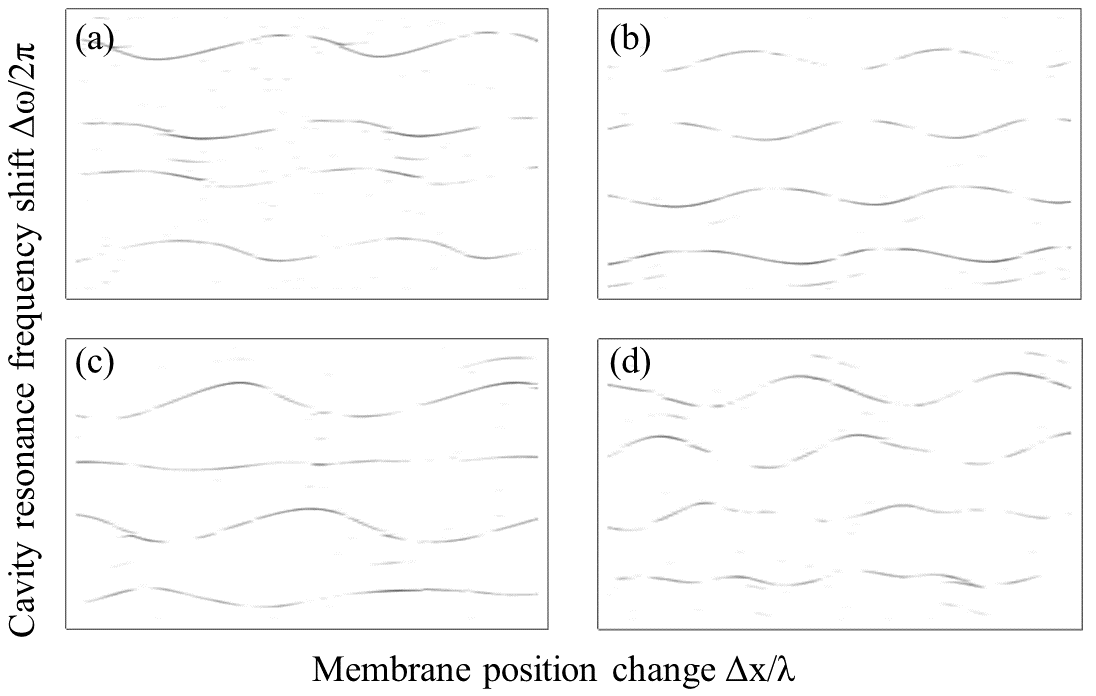}
\caption{Corresponding experimental measurements of Figs. 3(e)-3(h), respectively.}
\end{figure}

The experimental results of the asymmetric case are presented in Fig. 4. Figures 4(a)-4(d) are the corresponding experimental data of Figs. 3(e)-3(h), respectively. As one can see that Figs. 4(a)-4(d) excellently agree with the curves in the dashed squares of Figs. 3(e)-3(h).

\begin{figure}
\centering
\includegraphics[width=1\linewidth]{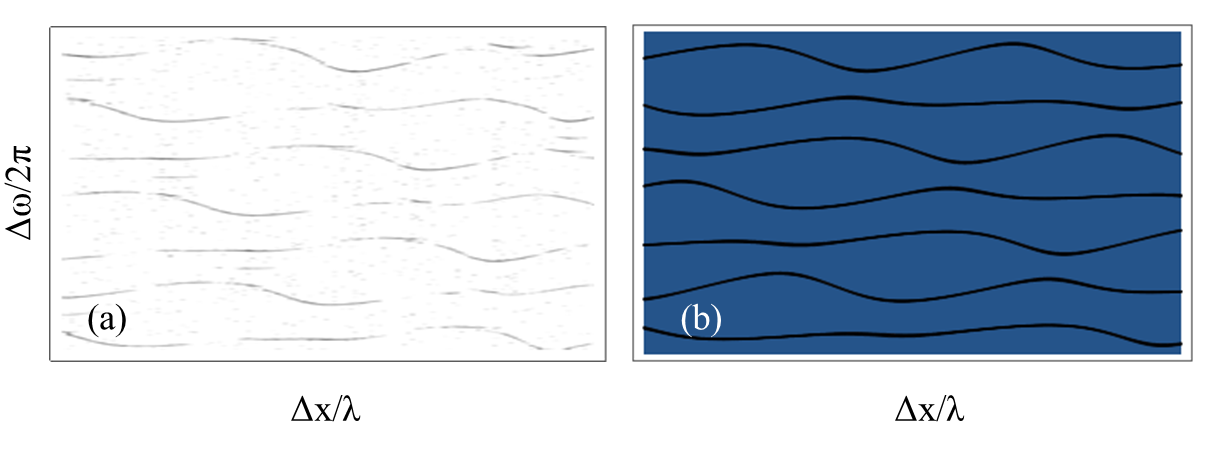}
\caption{(a) Experimental measurement and (b) theoretical simulation of the bands for the collective mechanical mode $\Delta x=(4\Delta {\rm{x}}_{\rm{1}}  + 3\Delta {\rm{x}}_{\rm{2}})/5$.}
\end{figure}

Actually, the collective mechanical mode can be any combinations of two membranes' motions, i.e. ${\Delta x=\rm{u}}\Delta {\rm{x}}_{\rm{1}}  + v\Delta {\rm{x}}_{\rm{2}}$ and $u$, $v$ can be any real numbers, which means that the projection of the 3D band-structure can be on a plane of an arbitrary angle with respect to $\Delta x_1$ axis, besides the four special cases analyzed above. Measured bands with $u/v=4/3$ are presented in Fig. 5(a). The corresponding theoretical simulation is shown in Fig. 5(b).

Lastly, we demonstrate that the eigenfrequencies of the vibrational modes of both membranes can be tuned individually with piezoelectricity. The eigenfrequencies of the vibrational modes are $f_{ij}  = \sqrt {\sigma (i^2  + j^2 )/4\rho l^2 }$, where $\sigma$ is the tensile stress, $\rho$ is the mass density, $l$ is the side length of the square membrane, and $i$, $j$ are the positive integer mode indices \cite{doi:10.1063/1.2884191,doi:10.1063/1.4862031,PhysRevLett.103.207204,PhysRevLett.112.127201,doi:10.1063/1.4967496,Tsaturyan2017}. Although the two membranes used in the experiment have very similar geometry, they are not identical. Slight difference in the side lengths leads to slightly different eigenfrequencies of vibrational modes. A tunable eigenfrequency is necessary for studying many physical phenomena, such as synchronization \cite{pikovsky2003synchronization,PhysRevLett.107.043603,PhysRevA.96.023805}. In the experiment, the membrane frame is directly glued to the ring piezo actuator \cite{doi:10.1063/1.5009952}. As the piezo actuator expands in the thickness mode, it will contact the radial mode. This leads to the deformation of the frame and the modification of membrane's stress and spring constant. This technique has a faster response compared to the photothermal effect \cite{doi:10.1063/1.3646914}.

\begin{figure}
\centering
\includegraphics[width=1\linewidth]{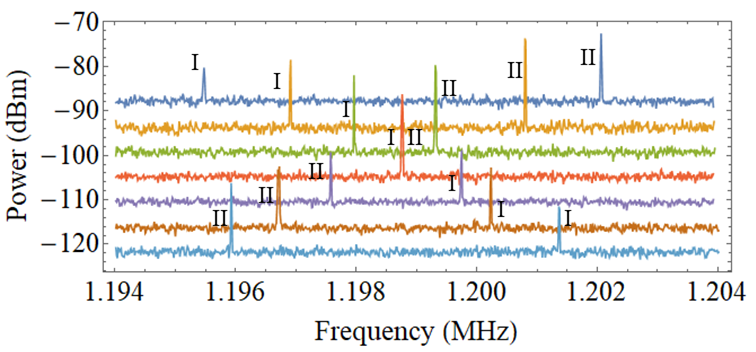}
\caption{Tunability of the mechanical vibrational modes.}
\end{figure}

Figure 6 shows the vibrational (3,3) modes of two membranes under Brownian motions. One can see that the frequencies of vibrational modes can be tuned, by changing the voltage applied on the piezos, where the membranes are glued on. The two modes can be either degenerated (the middle curve in Fig. 6) or separated as large as 7 $kHz$. The modes of two membranes are marked by I and II in Fig. 6. The curves in Fig. 6 are shifted vertically for clarity. The variation of signal strength during the tuning process is because the membranes' positions change, which leads to the modification of optomechanical coupling strength.  

In conclusion, we have realized a well controllable two-membrane-in-the-middle cavity optomechanical system. The band-structure-like diagrams are comprehensively analyzed in various cases, which is essential for further studies in such a two-membrane optomechanical system. These results provide a promising perspective for multimode optomechanics with distinct mechanical oscillators.

This research is supported by the National Key Research and Development Program of China (2017YFA0304201), National Natural Science Foundation of China (11374101, 91536112, 11704126), Natural Science Foundation of Shanghai (17ZR1443100), the Shanghai Sailing Program (17YF1403900), the Program for Professor of Special Appointment (Eastern Scholar) at Shanghai Institutions of Higher Learning.

\vskip 0.1in

$^*$jtsheng@lps.ecnu.edu.cn

$^\dag$hbwu@phy.ecun.edu.cn

\bibliographystyle{apsrev4-1}
%

\end{document}